\def\aa{{A\&A}}

\def\aj{{AJ}}

\def\apj{{ApJ}}

\def\mnras{{MNRAS}}

\documentclass[cup5b]{caps}
\usepackage{graphicx}
\usepackage{amssymb}
\usepackage{ociwsymp4}
\HeadText{E. K. Grebel}  

\begin{document}
\pagenumbering{arabic}
\setcounter{page}{237}

\author[]{EVA K. GREBEL\\Max-Planck Institute for Astronomy, 
Heidelberg, Germany\\Astronomical Institute of the University of
Basel, Switzerland}

\chapter{The Evolutionary History \\ of Local Group Irregular Galaxies}

\begin{abstract}

Irregular (Irr) and dwarf irregular (dIrr) galaxies are gas-rich galaxies with
recent or ongoing star formation.  In the absence of spiral density
waves, star formation occurs largely stochastically.  The scattered
star-forming regions tend to be long-lived and migrate slowly.  Older
populations have a spatially more extended and regular distribution.  In
fast-rotating Irrs high star formation rates with stronger concentration
toward the galaxies' center are observed, and cluster formation is facilitated.
In slowly or nonrotating dIrrs star formation regions are more widely
distributed, star formation occurs more quiescently, and the formation of 
OB associations is common.  On average, Irrs and dIrrs are 
experiencing continuous star formation with amplitude variations 
and can continue to form stars for another Hubble 
time.   

Irrs and dIrrs exhibit lower effective yields than spirals, and [$\alpha$/Fe]
ratios below the solar value.  This may be indicative of fewer 
Type II supernovae and lower astration rates in their past 
(supported by their low present-day star
formation rates).  Alternatively, many metals may be lost from the shallow
potential wells of these galaxies due to selective winds.  The differences
in the metallicity-luminosity relation between dIrrs and dwarf spheroidals
(which, despite their lower masses, tend to have too high a metallicity
for their luminosity as compared to dIrrs) lends further support to the idea 
of slow astration and slow enrichment in dIrrs.  The current data on
age-metallicity relations are still too sparse to distinguish between infall,
leaky-box, and closed-box models.  The preferred location of dIrrs in the
outer parts of galaxy groups and clusters and in the field as well
as the positive correlation between gas content and distance from massive
galaxies indicate that most of the dIrrs observed today probably have not yet 
experienced significant interactions or galaxy harassment.

\end{abstract}

\section{Introduction}\label{Sect_intro}

In this contribution, I will focus on the evolutionary histories 
of irregular (Irr) and dwarf irregular (dIrr) galaxies, including their 
chemical evolution.  
The name ``irregular'' refers to the irregular, amorphous appearance of
these galaxies at optical wavelengths, where the light contribution tends
to be dominated by scattered bright H\,{\sc ii} regions and their young,
massive stars.  Irrs are typically gas-rich galaxies that lack
spiral density waves as well as a discernible bulge or nucleus.  
Many Irrs are disk galaxies and appear to be an extension of late-type
spirals.  The most massive
disky Irrs with residual spiral structure are also called Magellanic
spirals; e.g., the Large Magellanic Cloud (LMC) (Kim et al.\ 1998) is a barred
Magellanic spiral.  Looser and more amorphous Irrs like the Small Magellanic
Cloud (SMC) are sometimes also
referred to as Magellanic irregulars (or barred Magellanic
irregulars if a bar is present); see de~Vaucouleurs (1957).  
A different system of subdivisions
was suggested by van~den~Bergh in his DDO luminosity classification
system (van~den~Bergh 1960, 1966).
DIrrs are simply less massive, less luminous Irrs; the distinction
between the two is a matter of definition rather than physics.  Typical
characteristics of dIrrs are a central surface brightness $\mu_V \lesssim 23$ 
mag arcsec$^{-2}$, a total mass of $M_{\rm tot} \lesssim 10^{10} M_{\odot}$, 
and an H\,{\sc i} mass of $M_{\rm HI} \lesssim 10^9 M_{\odot}$.
Solid body rotation is common among the Irrs and more massive dIrrs,
while low-mass dIrrs do not show measurable rotation; here random motions
dominate.  
A typical characteristic of Irrs and dIrrs alike is ongoing or recent
star formation.  The star formation intensity may range from burst-like,
strongly enhanced activity, to slow quiescent episodes.
Irrs and dIrrs can continue to form stars over a Hubble time
(Hunter 1997).

Substantial progress has been made in the photometric exploration of the 
star formation histories of Irrs and dIrrs over the past decade, largely
thanks to the superior resolution of the {\it Hubble Space Telescope}.
There is still very little known about the progress of the chemical
evolution in these galaxies as a function of time, but the advent of 
6-m to 10-m class telescopes and their powerful spectrographs is beginning 
to change this situation.  The most detailed information is available for
nearby Irr and dIrr galaxies in the Local Group, most notably the LMC, 
which is the Irr galaxy closest to the Milky Way.

\begin{figure*}
\centering
\includegraphics[width=0.85\columnwidth,angle=270]{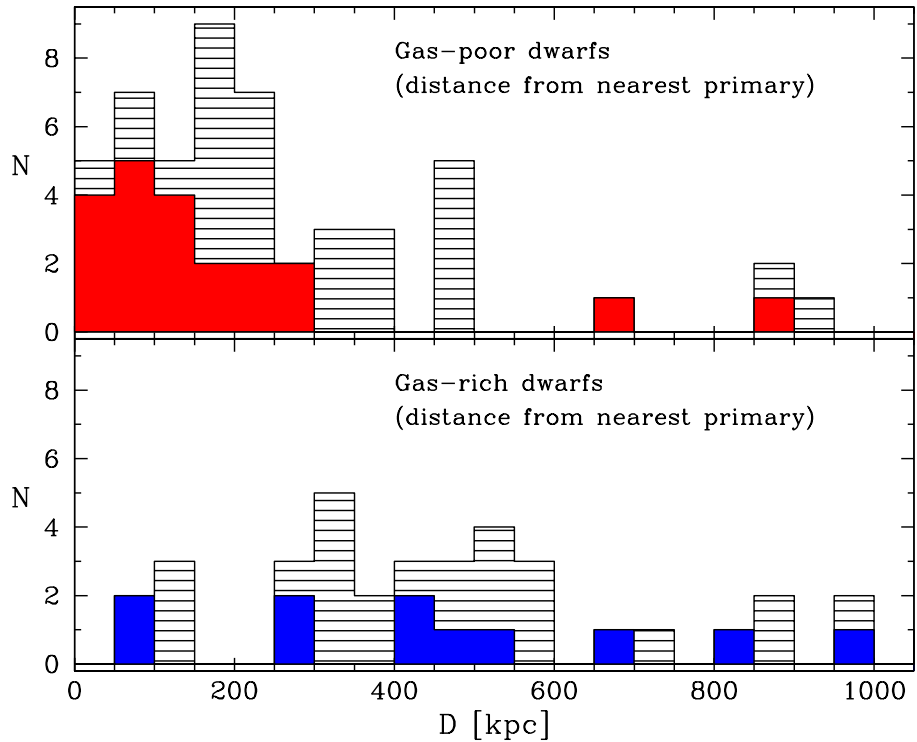}
\vskip 0pt \caption{
Morphological segregation in the Local Group (filled
histograms; see Grebel 2000) and in the M81 and Cen A groups (dashed 
histograms; input data from Karachentsev et al.\ 2002a, b).  Note 
the pronounced concentration of gas-poor, early-type dwarfs around
the nearest massive primary galaxy, while the gas-rich, late-type
dwarfs show less concentration and are more widely distributed.
This may be a signature of the impact of environmental effects, such
as gas stripping.}
\label{fig1}
\end{figure*}

\section{Distribution and Census of Irregulars in the Local Group}\label{Sect_census}

The Local Group, our immediate cosmic neighborhood, resembles other nearby
galaxy groups in many ways, including in its galaxy content, structure, mass,
and other properties (e.g., Karachentsev et al.\ 2002a, b).  It
is our best local laboratory to study galaxy evolution at the highest
possible resolution and in the greatest possible detail.  The Local Group
contains two dominant spiral galaxies surrounded by a large number of 
smaller galaxies.  Thirty-six galaxies are currently believed to be members
of the Local Group if a zero-velocity surface of 1.2 Mpc is adopted 
(Courteau \& van~den~Bergh 1999; Grebel, Gallagher, \& 
Harbeck 2003)\footnote{Note that recent kinematic estimates suggest an
even smaller radius of ($0.94\pm 0.10$) Mpc for the zero-velocity surface
(Karachentsev et al.\ 2002c), which reduces the above number of Local
Group dwarf galaxies by two.}.  
The smaller galaxies in
the Local Group include a spiral galaxy (M33), 11 gas-rich Irr 
and dIrr galaxies (including low-mass, so-called transition-type
galaxies that comprise properties of both dIrrs and dwarf spheroidals), 
four elliptical and dwarf
elliptical galaxies, and 17 gas-deficient dwarf spheroidal (dSph) galaxies.
For a listing of the basic 
properties of these galaxies, see Grebel et al. (2003).      
Their three-dimensional distribution is illustrated in Grebel (1999; Fig.\
3). 
Recent reviews of Local Group galaxies include Grebel (1997, 1999, 2000),
Mateo (1998), and van~den~Bergh (1999, 2000).  

DIrrs are the 
second most numerous galaxy type in the Local Group.  While new dwarf
members of the Local Group are still being discovered (e.g., Whiting,
Hau, \& Irwin 1999), these tend to be gas-deficient, low-mass dSph
galaxies, which have intrinsically low optical surface brightnesses
and cannot be found from their H\,{\sc i} 21~cm emission lines.  
The Irr and dIrr census of the Local Group appears to be 
complete.  

Irrs and dIrrs are found in galaxy groups and clusters as
well as in the field and exhibit little concentration toward massive
galaxies in contrast to early-type dwarfs.  This 
morphological segregation is clearly seen in the Local Group and in
nearby groups (Fig.\ \ref{fig1}).  It becomes even more pronounced in galaxy
clusters, where the distribution of Irrs shows the least concentration
of all galaxy types toward the cluster core (e.g., Conselice, Gallagher,
\& Wyse 2001, and references therein), which has been attributed to 
continuing infall of Irrs and subsequent harassment.  Conversely, in
very loose groups or ``clouds'' (such as the Canes Venatici I Cloud)
that are still far from approaching  
dynamical equilibrium, an overabundance of Irrs and dIrrs is observed
as compared to early-type dwarfs (Karachentsev et al.\ 2003a), indicative 
of a lack of interactions.

\section{The Interstellar Medium of Local Group Irregulars}\label{Sect_ISM}

\subsection{The Magellanic Clouds}\label{Sect_ISM_MCs}

The Magellanic Clouds are the two most massive Irrs in the Local
Group, and the only two Irrs in immediate proximity to a 
massive spiral galaxy.  Their distances from the Milky Way are
50 kpc (LMC) and 60 kpc (SMC), respectively.
They are the only two Local Group Irrs that are closely
interacting with each other (and with the Milky Way).  
According to the earlier definition, 
the SMC qualifies as a dIrr.  

The global distribution of neutral hydrogen within the Magellanic Clouds and
other comparatively massive Irrs tends to show a 
regular, symmetric appearance, in contrast to their visual morphology.
On smaller scales, the H\,{\sc i} is flocculent and 
exhibits a complicated fractal pattern full of shells and clumps
(e.g., Kim et al.\ 1998; Stanimirovic et al.\ 1999).  The lack of
correlation between the H\,{\sc i} shells and the optically dominant
H\,{\sc ii} shells suggests that H\,{\sc i} shells live longer than
the OB stars that caused them initially (Kim et al.\ 1999).  The 
H\,{\sc i} associated with H\,{\sc ii} regions is usually more extended
than the ionized regions.  The fractal structure of the neutral gas 
is self-similar on scales from tens to hundreds of pc
(Elmegreen, Kim, \& Staveley-Smith 2001) and appears to result from
the turbulent energy input caused by winds of recently formed massive stars
and supernova explosions.  

With $0.5 \times 10^9$ $M_{\odot}$ (Kim et al.\ 1998) the LMC's gaseous
component contributes about 9\% to its total mass, while 
it is $\sim 21$\% in the SMC (H\,{\sc i} mass of $4.2\times10^8$ $M_{\odot}$;
Stanimirovic et al.\ 1999).  In comparison to the Milky Way, the 
gas-to-dust ratio is roughly 4 times lower in the LMC (Koornneef 1982)
and about 30 times lower in the SMC (Stanimirovic et al.\ 2000), implying 
a smaller grain surface area per hydrogen atom, fewer coolants, 
and thus a reduced H$_2$ formation efficiency (Dickey et al.\ 2000;
Stanimirovic et al.\ 2000;
Tumlinson et al.\ 2002).  Indeed, the total diffuse H$_2$ mass is only
$8\times 10^6$ $M_{\odot}$ in the LMC and $2\times 10^6$ $M_{\odot}$ in the
SMC, which corresponds to 2\% and 0.5\% of their H\,{\sc i} masses,
respectively (Tumlinson et al.\ 2002).  Also the reduced CO emission
from both Clouds (3--5 times lower than expected for Galactic giant
molecular clouds) is indicative of the high UV radiation field in 
low-metallicity environments and hence high CO photodissociation rates
(Israel et al.\ 1986; Rubio, Lequeux, \& Boulanger 1993).
While high dust content is correlated with high H$_2$ concentrations,
H$_2$ does not necessarily trace CO or dust (Tumlinson et al.\ 2002).

Photoionization through massive stars
is the main contributor to the optical appearance of
the interstellar medium (ISM) at $\sim 10^4$ K
in the Clouds and other gas-rich, star-forming galaxies.
The LMC has a total H$\alpha$ luminosity of $2.7\times10^{40}$ erg s$^{-1}$;
30\% to 40\% is contributed by diffuse, extended gas
(Kennicutt et al.\ 1995).  In the LMC 
nine H\,{\sc ii} supershells with diameters $>600$
pc are known 
(Meaburn 1980).  Their rims are marked by strings of H\,{\sc ii} regions
and young clusters/OB associations.  The standard picture for supershells
suggests that these are expanding shells driven by propagating star
formation (e.g., McCray \& Kafatos 1987).  However, an age
{\em gradient} consistent with this scenario was not
detected in the largest of these supershells, LMC4 (Dolphin \&
Hunter 1998).  Nor are other LMC supershells
expanding as a whole, but instead appear to consist of hot gas confined
between H\,{\sc i} sheets and show localized expansion.
Supershells in several other galaxies neither
show evidence for expansion (e.g., Points et al.\ 1999), nor 
the expected young massive stellar populations (Rhode et al.\ 1999).
In contrast, the three H\,{\sc i} supershells and 495 giant shells
in the SMC appear to be expanding (Staveley-Smith et al.\ 1997;
Stanimirovic et al.\ 1999).

The hot, highly ionized corona of the LMC with 
collisionally ionized gas (temperatures $\gtrsim 10^5$ K) (Wakker et al.\ 1998)
is spatially uncorrelated with star-forming regions.  
A hot halo is also observed around the SMC,
but here clear correlations with star-forming 
regions are seen.  This corona may be caused in part by gas falling
back from a galactic (i.e., SMC) fountain (Hoopes et al.\ 2002).  
The O\,{\sc vi} column density exceeds the corresponding Galactic 
value by 1.4 (Hoopes et al.\ 2002), consistent with the longer cooling
times expected at lower metallicities (Edgar \& Chevalier 1986).

The Magellanic Clouds, which only have a deprojected 
distance of 20 kpc from each
other, interact with each other and with the Milky Way.  Apart from
an impact on the structure and star formation histories of these three
galaxies (e.g., Hatzidimitriou, Cannon, \& Hawkins 1993;
Kunkel, Demers, \& Irwin 2000; Weinberg 2000; van~der~Marel et al.\ 2002),
this has given rise to extended gaseous
features surrounding the Magellanic Clouds (Putman et al.\ 2003, and 
references therein).  
Part of these are likely caused by tidal interactions, but ram pressure
appears to have played an important role as well (Putman et al.\ 1998;
Mastropietro et al.\ 2004).  Metallicity determinations for gas in the
Magellanic Stream, which is trailing behind the Magellanic Clouds and subtends
at least $10^{\circ} \times 100^{\circ}$ on the sky, confirm that the gas 
is not primordial (Lu et al.\ 1998;
Gibson et al.\ 2000).  The  H$_2$ detected in the leading arm of the 
Stream may originally have formed in the SMC (Sembach et al.\ 2001).
No stars are known to be connected with the 
Magellanic Stream (Putman et al.\ 2003).

Another prominent H\,{\sc i} feature is the ``Magellanic Bridge''
or InterCloud region
($10^8$ $M_{\odot}$; Putman et al.\ 1998), which connects the LMC 
and SMC.  Cold (20 to 40 K) H\,{\sc i} gas has been detected in the Bridge
(Kobulnicky \& Dickey 1999), and recent star formation
occurred there over the past 10 to 25 Myr (Demers \& Battinelli 1998).
Intermediate-age stars are also present in parts of the Bridge (carbon
stars: Kunkel et al.\ 2000, and references therein).
Higher ionized species with temperatures up to $\sim 10^5$
K show an abundance pattern suggesting depletion into dust (Lehner et al.\ 
2000).  Interestingly, the metallicities of young stars in the Bridge
were found to be [Fe/H] $\approx\, -1.1$ dex (Rolleston et al.\ 1999),
0.4 dex below the mean abundance of the young SMC population, which is
inconsistent with the proposed tidal origin 200 Myr ago (Murai \& 
Fujimoto 1980; Gardiner \& Noguchi 1996).

\subsection{More Distant Dwarf Irregular Galaxies}\label{Sect_ISM_dwarfs}

The other Local Group dIrrs are more distant
from the dominant spirals, and fairly isolated.  Interactions may 
still occur, but if this happens the interaction partners tend to be gas 
clouds rather than galaxies.
Generally, star formation activity and gas content decrease with galaxy 
mass, but
the detailed star formation histories and ISM properties of the dIrrs
present a less homogeneous picture. 

NGC\,6822, a dIrr at a distance of $\sim 500$ kpc, is 
embedded in an  elongated H\,{\sc i} cloud with numerous shells and holes.
Its
total  H\,{\sc i} mass is $1.1 \times 10^8$ $M_{\odot}$, $\sim 7$\% of its total
mass.  The masses of individual CO clouds reach up to $(1-2)\times10^5$
$M_{\odot}$ (Petitpas \& Wilson 1998), while the estimated H$_2$ content 
is 15\% of the H\,{\sc i} mass (Israel 1997), and the dust-to-gas mass ratio is
$\sim 1.4 \times 10^{-4}$ (Israel, Bontekoe, \& Kester 1996).  
NGC\,6822 contains many H\,{\sc ii} regions.  Its
huge supershell ($2.0\times 1.4$ kpc) was likely caused by the passage of
and interaction with a nearby $10^7$  $M_{\odot}$ H\,{\sc i} cloud
and does not show signs of expansion (de~Blok \& Walter 2000).
The older stars in IC\,10 describe an elliptical, extended halo (Letarte
et al.\ 2002) distinct from the elongated H\,{\sc i} distribution.  The 
latter, however, is traced closely by a population of young blue stars ($\sim
180$ Myr) that appear to have formed following the interaction with the 
passing H\,{\sc i} cloud (de~Blok \& Walter 2003; Komiyama et al.\ 2003) some 300 Myr ago.  In NGC\,6822, the H\,{\sc i} distribution is thus
only slightly more extended than the stellar loci.

The H\,{\sc i} of IC\,10 (distance 660 kpc) is 7.2 times more extended than
its Holmberg radius (Tomita, Ohta, \& Sait\={o} 1993).  While the inner 
part of the neutral hydrogen of IC\,10 is a regularly rotating disk full of 
shells and holes, the outer H\,{\sc i} gas is counter-rotating (Wilcots \& 
Miller 1998).
IC\,10 is currently experiencing a massive starburst, which is possibly
triggered and fueled by an infalling H\,{\sc i} cloud 
(Sait\={o} et al.\ 1992; Wilcots \& Miller 1998).
IC\,10 contains a nonthermal superbubble that may be the
result of several supernova explosions (Yang \& Skillman 1993).
The masses of the CO clouds in IC\,10 appear to be 
as high as up to $5\times10^6$ $M_\odot$ 
(Petitpas \& Wilson 1998), which would indicate that more than 20\% of this
galaxy's gas mass is molecular.  Owing to the high radiation field and the
destruction of small dust grains, the ratio of far-infrared
[C\,{\sc ii}] to CO 1--0 emission is a factor 4 larger than in the 
Milky Way (Bolatto et al.\ 2000), resulting in small CO
cores surrounded by large [C\,{\sc ii}]-emitting envelopes 
(Madden et al.\ 1997).
Two H$_2$O masers were detected in dense clouds in IC\,10, marking sites of
massive star formation (Becker et al.\ 1993).  The internal dust content of 
IC\,10 is high, and its properties prompted Richer et al.\ (2001) to suggest
that this galaxy should actually be classified as a blue compact dwarf.

Less detailed information is available for the ISM in the other Local Group
dIrrs, which do not appear to be involved in ongoing interactions and which
are evolving fairly quiescently.  The H\,{\sc i} in these dIrrs may be up 
to 3 times more extended than the optical galaxy and is clumpy on scales 
of 100 to 300 pc.  The most massive clumps reach $\sim 10^6$
$M_{\odot}$.  H\,{\sc i} concentrations tend to be close to
H\,{\sc ii} regions.  Some dIrrs contain cold H\,{\sc i} clouds
associated with molecular gas.
The total H\,{\sc i} masses are usually $< 10^9$
$M_{\odot}$, and less than $10^7$ $M_{\odot}$ for transition-type dwarfs.  
The center of the H\,{\sc i} distribution coincides
roughly with the optical center of the dIrrs, although the H\,{\sc i} may
show a central depression surrounded by an H\,{\sc i} ring or arc (e.g.,
SagDIG, Leo\,A), possibly a consequence of star formation, or the
H\,{\sc i} may be off-centered (e.g., Phoenix; St-Germain et al.\ 1999).
In low-mass dIrrs there are no signatures of rotation, but these may be
obscured by expanding shells and bubbles.
Further details are given in Lo, Sargent, \& Young (1993), Young \& Lo (1996, 
1997), Elmegreen \& Hunter (2000), and Young et al.\ (2003).  

Lower gravitational
pull and the lack of shear in the absence of differential rotation imply
that H\,{\sc i} shells may become larger and are long-lived
(Hunter 1997).  Diameters, ages, and expansion velocities of the
H\,{\sc i} shells increase with later Hubble type (Walter \& Brinks 1999)
and scale approximately with the square root of the galaxy luminosity
(Elmegreen et al.\ 1996).  Shell-like
structures, H\,{\sc i} holes, or off-centered gas may be driven by
supernovae and winds from massive stars
following recent star formation episodes or tidal interactions.

For a review on nebular abundances in Irrs, see the contribution by 
Garnett (2004).  Here it should only be mentioned
that the effective yields in Irrs computed from gas-phase abundances 
are lower than those in the main stellar disks of
spirals.  Lower effective yields are also correlated with lower 
rotational velocities (Garnett 2002).  This is interpreted as preferential 
metal loss
through winds in the more shallow potential wells of Irrs and dIrrs,
but may also be due to lower astration levels (e.g., Pilyugin \& 
Ferrini 2000).    For a review of the general ISM properties in Local 
Group dwarf galaxies, see Grebel (2002a).

\section{Large-scale Star Formation and Spatial Variations}\label{Sect_SF}

The dwarf galaxies in the Local Group vary widely in their
star formation and enrichment histories,
times and duration of their major star formation episodes, 
and fractional distribution of ages and subpopulations.
Indeed, when studied in detail, no two dwarf galaxies turn out to be
alike, not even if they are of the same morphological type or have similar
luminosities (Grebel 1997).  On the other hand, in spite of their 
individual differences, they do follow certain common global 
correlations such as increasing mean metallicity with luminosity 
(\S \ref{Sect_global_LZ}).  

%\begin{figure*}
%\centering
%\includegraphics[width=10.0cm,angle=0]{grebel_fig2.ps}
%\vskip 0pt \caption{
%Large-scale star formation patterns in the Large Magellanic
%Cloud spanning the past $\sim 250$ Myr.  The individual dots correspond
%to age-dated Cepheids ({\it upper panel}) and supergiants ({\it lower panel}). 
%A few prominent features like the LMC bar, supershell LMC\,4, and
%30\,Doradus are marked by solid and dashed lines.
%Note how star formation migrated along the LMC's bar and finally 
%vanished in its southernmost past, and how other regions such as 
%30\,Doradus and LMC\,4 only became strongly active over the past $\sim$
%30 Myr.  Within the time scales depicted here, which incidentally
%correspond to roughly one rotation period, stars are not expected to
%have migrated far from their birthplaces. (From Grebel \& Brandner 1998.) 
%}
%\label{fig2}
%\end{figure*}

\subsection{Large-scale Star Formation}\label{Sect_SF_scale}

The ISM properties of Irrs and dIrrs outlined in the previous section
already show that there are spatial variations in star formation history and
other characteristics within these galaxies.  In general, dwarf galaxies
of all types show a tendency for the younger
populations to be more centrally concentrated (and possibly more 
chemically enriched),
whereas older populations are more extended (Grebel 1999, 2000; Harbeck
et al. 2001).  In Irr and dIrr galaxies, H\,{\sc ii} regions tend to be
located within the part of the galaxy that shows solid body rotation and
are usually even more centrally concentrated (Roye \& Hunter 2000).  
Star-forming regions may, however, be found out to six optical scale
lengths, indicating that star formation is truncated at lower gas
density thresholds than in spirals (Parodi \& Binggeli 2003).  
In dIrrs dominated by chaotic motions, the degree of central concentration
of recent star formation is lower, whereas fast-rotating Irrs tend to
exhibit the highest central concentrations.  The same trend also holds 
for the star formation activity: low-mass dIrrs with no measurable
rotation also have lower star formation rates (Roye \& Hunter 2000;
Parodi \& Binggeli 2003).

\begin{figure*}
\centering
\includegraphics[width=10.0cm,angle=0]{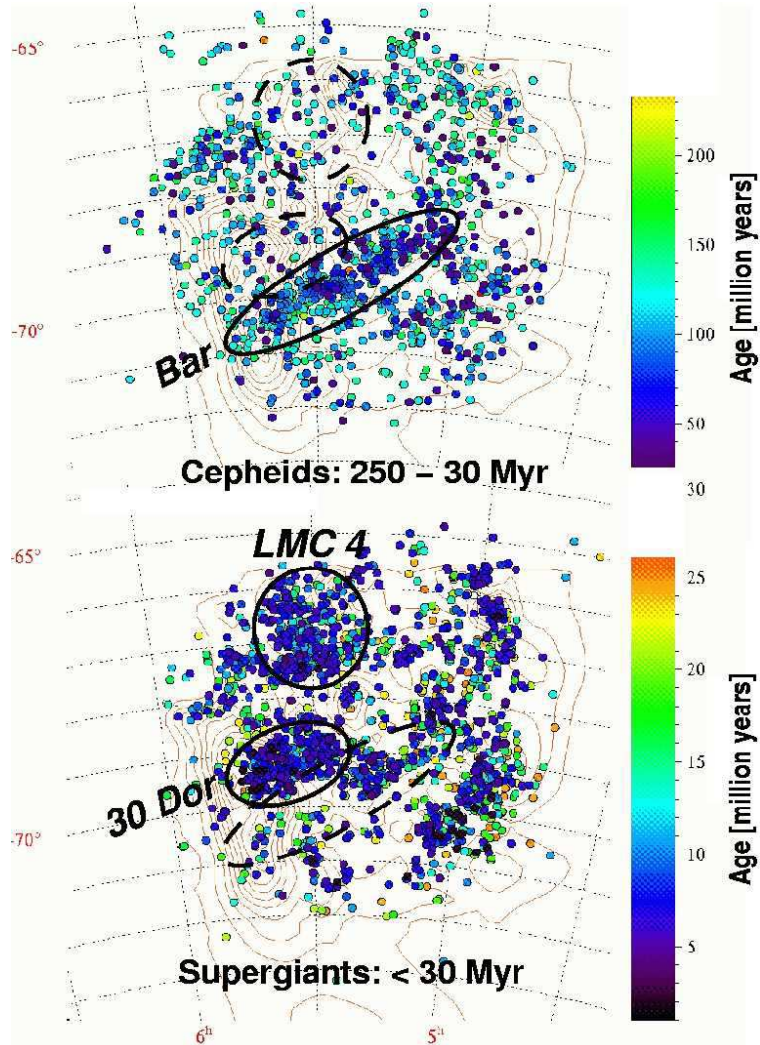}
\vskip 0pt \caption{
Large-scale star formation patterns in the Large Magellanic
Cloud spanning the past $\sim 250$ Myr.  The individual dots correspond
to age-dated Cepheids ({\it upper panel}) and supergiants ({\it lower panel}).
A few prominent features like the LMC bar, supershell LMC\,4, and
30\,Doradus are marked by solid and dashed lines.
Note how star formation migrated along the LMC's bar and finally
vanished in its southernmost past, and how other regions such as
30\,Doradus and LMC\,4 only became strongly active over the past $\sim$
30 Myr.  Within the time scales depicted here, which incidentally
correspond to roughly one rotation period, stars are not expected to
have migrated far from their birthplaces. (From Grebel \& Brandner 1998.)
}
\label{fig2}
\end{figure*}

How does star formation progress in irregular galaxies?
Irrs and the more massive dIrrs usually contain multiple distinct regions
of concurrent star formation.
These regions often remain active for several 100 Myr, are found 
throughout the
main body of these galaxies (see above), and can migrate.  This is 
illustrated in Figure\ \ref{fig2}
for the LMC, where the large-scale star formation history of the last
$\sim 250$ Myr (approximately one rotation period) is shown
(see Grebel \& Brandner 1998 for full details).  Note how some of the
active regions have continued to form stars over extended periods 
and propagated slowly, 
whereas others only became active during the past 30 Myr.
The star formation complexes resemble superassociations and may span 
areas of a few hundred pc (Grebel \& Brandner 1998).  
In supershells, typical time scales for continuing star formation 
on length scales of 0.5 kpc range from 15 to 30 Myr, usually without showing
clear signs of spatially directed propagation with time 
(see also Grebel \& Chu 2000 and \S \ref{Sect_ISM_MCs}).
CO shows a strong correlation
with H\,{\sc ii} regions and young ($<10$ Myr) clusters,
but only little with older clusters and supernova remnants (Fukui et al.\ 1999; cf.\ Banas et al.\ 1997).
Massive CO clouds have
typical lifetimes of $\sim 6$ Myr and are dissipated within $\sim$3 Myr
after the formation of young clusters (Fukui et al.\ 1999; Yamaguchi et
al.\ 2001).
Spatially resolved star formation histories have also been derived for
two dIrr galaxies just beyond the Local Group covering the past 500--700 Myr.
They reveal similar long-lived, gradually migrating zones of star formation
(Sextans A: Van~Dyk, Puche, \& Wong 1998; Dohm-Palmer et al.\ 2002;
GR\,8: Dohm-Palmer et al.\ 1998), as seen in the more massive Magellanic
Clouds.   

In low-mass dIrrs one usually observes only one single low-intensity 
star-forming region.  DIrrs and
transition-type dIrr/dSph galaxies tend to be fairly 
quiescent, often having experienced the bulk of their star formation
at earlier times.  (In fact, transition-type dwarfs resemble dSphs
in their gradually declining star formation rates; see Grebel et al.\
2003 for details.)
Evidence for migrating star formation is found in low-mass dIrrs as 
well, albeit on smaller scales owing to the smaller sizes of these
galaxies (e.g., Phoenix: Mart\'{\i}nez-Delgado, Gallart, \& Aparicio 1999).

\subsection{Intermediate-age and Old Stellar Populations}\label{Sect_SF_old}

Irrs and massive dIrrs tend to show extended halos of intermediate-age
stars (ages $\sim 1$ to $\sim 10$ Gyr), which can be conveniently traced 
by carbon stars (e.g., Letarte et al.\ 2002).  In
the Magellanic Clouds, the density distributions of different populations
ages become increasingly more regular and extended with increasing age
(e.g., Cioni et al.\ 2000; Zaritsky et al.\ 2000), whereas the young
populations are responsible for the irregular appearance of these two
galaxies.  The centroids of the different populations do not always
coincide.  Features resembling stellar bars are found in many dIrrs, which
do not necessarily coincide with the peak H\,{\sc i} distribution or its
centroid.
In low-mass dIrrs there are not enough intermediate-age
tracers such as C stars to say much about the distribution of these
populations (see, e.g., Battinelli \& Demers 2000); the number of C
stars decreases with absolute galaxy luminosity and also with galaxy
metallicity (see Groenewegen 2002 for a recent review and census of
C stars in the Local Group).

All Irr and dIrr galaxies examined in detail so far show clear evidence for 
the presence of old ($> 10$ Gyr)
populations, a property that they appear to share
with all galaxies whose stellar population have been resolved.
For instance, deep ground-based imaging of the ``halos'' of dIrrs led to
the detection of old red giant branches (e.g., Minniti \& Zijlstra 1996;
Minniti, Zijlstra, \& Alonso 1999).  In closer dIrrs, horizontal branch stars
have been detected in field populations (e.g., IC 1613: Cole et al.\ 1999;
Phoenix: Holtzman, 
Smith, \& Grillmair 2000; WLM: Rejkuba et al.\ 2000; Leo\,A: Dolphin et
al.\ 2002) and in globular
clusters (e.g., WLM: Hodge et al.\ 1999).  Horizontal branch stars are
unambiguous tracers of ancient populations.  In the 
closest dIrrs and Irrs even the old main sequence turn-offs have been
resolved, allowing differential age dating.   Interestingly (with the
possible exception of the SMC), the  oldest datable populations in
all nearby dwarf galaxies 
turn out to be indistinguishable
in age from each other and from the Milky Way,
indicating a common epoch of early star formation (e.g., Olsen et al.\
1998; Johnson et al.\ 1999; see Grebel 2000 for 
a full list of references).  Apart from the recent interaction
in NGC\,6822 (\S \ref{Sect_ISM_dwarfs}), 
the old populations also are usually the most
extended ones. However, their fractions vary:  in some cases they only
constitute a tiny portion of the stellar content of their parent galaxy. 

\subsection{Modes of Star Formation}\label{Sect_SF_modes}

The ISM in dIrrs is
highly inhomogeneous and porous, full of small and large shells and
holes.  The global gas density tends to be
significantly below the Toomre criterion for star formation
(van~Zee et al.\ 1997).
Stochastic, star formation may be driven by homogeneous turbulence, which
creates local densities above the star formation threshold (e.g.,
Stanimirovic et al.\ 1999).  
Self-propagating stochastic star formation (Gerola \& Seiden 1978;
Gerola, Seiden, \& Schulman 1980; Feitzinger et al.\ 1981) 
can lead to structures of sizes of
up to 1 kpc, in which star formation processes remain active for 30--50
Myr, or to the formation of long-lived spiral features if an off-centered
bar is present (Gardiner, Turfus, \& Putman 1998).  In the absence of shear,
star formation continues along regions of high H\,{\sc i} column density,
fueled by the winds of recently formed stars and supernovae explosions.

Dense gas concentrations may, however, also remain inactive for hundreds
of Myr, and there are not usually obvious triggers for the onset of star
formation (see Dohm-Palmer et al.\ 2002).  This may be different in 
nonquiescently evolving, starbursting dIrrs like IC\,10:  gas
accretion or other interactions
may be triggering the starburst (see \S \ref{Sect_ISM_dwarfs}).  
The existence of
isolated dIrrs with continuous star formation outside of groups
shows that external triggers are not needed.  
Quiescently evolving dIrrs exhibit widely distributed star formation
and have very small color gradients, whereas starbursting dIrrs 
show much more concentrated star formation and strong color gradients
(van~Zee 2001).  
The analysis of 72 dIrr
galaxies in nearby groups and in the field
revealed that the radial distribution of star-forming regions
follows on average an annulus-integrated exponential distribution, and
that secondary star-forming peaks at larger distances are consistent
with internal triggering via stochastic, self-propagating star formation
(Parodi \& Binggeli 2003).

Quiescent dIrrs tend to form OB associations, while massive starbursts
can lead to the formation of more compact star clusters.  The 
number of massive clusters tends to correlate with galaxy mass (i.e.,
roughly with luminosity; see, e.g., Parodi \& Binggeli 2003).  For instance,
in the dIrr NGC\,6822 on average one cluster is formed per $6\times 10^6$
years (a much smaller number 
than in the more massive LMC), while an OB association forms every
$7\times 10^5$ years, similar to the LMC (Hodge 1980).  The distinctive,
well-separated peaks in the formation rate of populous
clusters in the LMC, however, seem to
be caused by close encounters with the Milky Way and the SMC (Gardiner, 
Sawa, \& Fujimoto 1994; Girardi et al. 1995; Lin, Jones, \& Klemola 1995);
it is surprising that no corresponding enhancement in the SMC's fairly
continuous cluster formation rate is seen.  Generally, old globular
clusters are rare in dIrrs.  For the cluster census in Local Group Irr and 
dIrr galaxies, see Table~\ref{table1}.

On a global, long-term scale, star formation in dIrrs has
essentially occurred continuously at a constant rate with amplitude
variations of 2--3 (Tosi et al.\ 1991; Greggio et al.\ 1993),
is largely governed by internal, local processes, 
and will likely continue for another Hubble time (Hunter 1997; van~Zee 2001).

%\begin{center}
\begin{table*}
\caption{Star Clusters in the Local Group Irr and dIrr Galaxies} 
\begin{tabular}{llrcrccr}
\hline \hline
Galaxy    & Type        &  D$_{\rm Sp}$ &  $M_V$ & $N_{GC}$    & $S_N$ &  [Fe/H]         & $N_{OC}$      \\
          &             &  [kpc]        &  [mag] &             &       &  [dex]          &               \\
	  \hline
	  LMC       & Ir\,{\sc iii-iv}   &   50 & $-18.5$& $\sim$13  & 0.5        &  $-2.3$, $-1.2$ & $\gtrsim$4000 \\
	  SMC       & Ir\,{\sc iv/iv-v}  &   63 & $-17.1$& 1         & 0.1        &  $-1.4$           & $\gtrsim$2000 \\
	  NGC\,6822 & Ir\,{\sc iv-v}     &  500 & $-16.0$& 1         & 0.4        &  $-2.0$           & $\sim$20 \\
	  WLM       & Ir\,{\sc iv-v}     &  840 & $-14.4$& 1         & 1.7        &  $-1.5$           & $\ge$1  \\
	  IC\,10    & Ir\,{\sc iv}:      &  250:& $-16.3$& 0         & 0          &   ---             & $\gtrsim$13 \\
	  IC\,1613  & Ir\,{\sc v}        &  500 & $-15.3$& 0         & 0          &  ---              & $\gtrsim$5  \\
	  Phe       & dIrr/dSph          &  405 & $-12.3$& [4:]       & [48:]     &  ---              & ?       \\
	  PegDIG    & Ir\,{\sc v}        &  410 & $-11.5$& 0         & 0          &  ---              & $\lesssim$3 \\
	  LGS\,3    & dIrr/dSph          &  280 & $-10.5$& 0         & 0          &  ---              & $\lesssim$13 \\
\hline \hline 
\end{tabular}
\footnotesize
Notes:\, Only galaxies known to contain star clusters are listed.
D$_{\rm Sp}$ denotes
the distance to the nearest spiral galaxy (M31 or Milky Way, Col.\ 3).  $N_{GC}$
and $N_{OC}$ (Cols.\ 5 \& 8) list the number of globular clusters
and open clusters, respectively.
Note that the globular cluster suspects in Phoenix are highly uncertain.
$S_N$ (Col.\ 6) is the specific globular cluster frequency.  When two
values are listed in Col.\ 7 (metallicity), these indicate the most
metal-rich and most metal-poor globular clusters.
For more details, a full list of galaxies with star clusters in the
Local Group, and references,
see Grebel (2002b).
\label{table1}
\end{table*}
%\end{center}

\section{Metallicity and Age}\label{Sect_AMR}

\subsection{Young Populations and Chemical Homogeneity}\label{Sect_AMR_young}

The gas in Irrs and dIrrs is fairly well mixed, and mixing must proceed
rapidly considering how homogeneous present-day
H\,{\sc ii} region abundances at
different locations within the same galaxy are.  Nebular abundances of
ionized gas trace the youngest populations and the chemical composition
of the star-forming material.  Why there is such a high degree of 
homogeneity is not fully understood, nor is it clear how the mixing
proceeds; mechanisms may include 
winds and turbulence.  Inhomogeneities are only 
expected to be detectable very shortly after the responsible stellar
population formed (e.g., when pollution by Wolf-Rayet stars occurs; 
see Kobulnicky
et al.\ 1997).  Note that such global chemical homogeneity appears to
be less pronounced in gas-deficient dSph galaxies, which seem to have 
experienced different star formation and chemical enrichment histories
than dIrrs (see, e.g., Harbeck et al.\ 2001; Grebel et al.\ 2003).  

If Irrs and dIrrs are chemically homogeneous,
one would expect to measure comparable abundances in H\,{\sc ii}
regions and young stars of a given dIrr, since both trace the same population.  
Due to their proximity, in the Magellanic Clouds supergiants, giants, and
even massive main sequence stars can be analyzed with high-dispersion
spectroscopy and individual element ratios can be measured.  
Indeed, in the Magellanic Clouds good agreement
is found between nebular and stellar abundances (e.g., Hill, Andrievsky,
\& Spite 1995;
Andrievsky et al.\ 2001).
Star-to-star variations in the overall metallicity of young stars
(B to K supergiants and B main sequence stars)
are small ($\pm 0.1$ dex: Hill 1997; Luck et al.\
1998; Venn 1999; Rolleston, Trundle, \& Dufton 2002), 
and there is no evidence for a significant Population I metallicity
spread in either Cloud.  Also,
the differences between the stellar abundances in young clusters and 
field stars are small (Gonzalez \& Wallerstein 1999; Hill 1999;
Korn et al.\ 2000, 2002; Rolleston et al.\ 2002).  
The mean metallicity of the young population
in the LMC is [Fe/H] $\approx\, -0.3$ dex and $\sim -0.7$ dex in the SMC.

Very good agreement between young stellar and nebular abundances 
is also found in the more distant dIrrs 
NGC\,6822 (B supergiants: Muschielok et al.\ 1999; A supergiants:
Venn et al.\ 2001, both yielding [Fe/H] = $-0.5$ dex), 
GR\,8, and Sex\,A (Venn et al. 2004).
The two blue supergiants analyzed in WLM, on the other hand, have 
clearly higher metallicities than found in its H\,{\sc ii} regions
(Venn et al.\ 2003).  The reasons for this discrepancy in WLM are still
unknown.

\begin{figure*}
\centering
\includegraphics[width=1.00\columnwidth,angle=0]{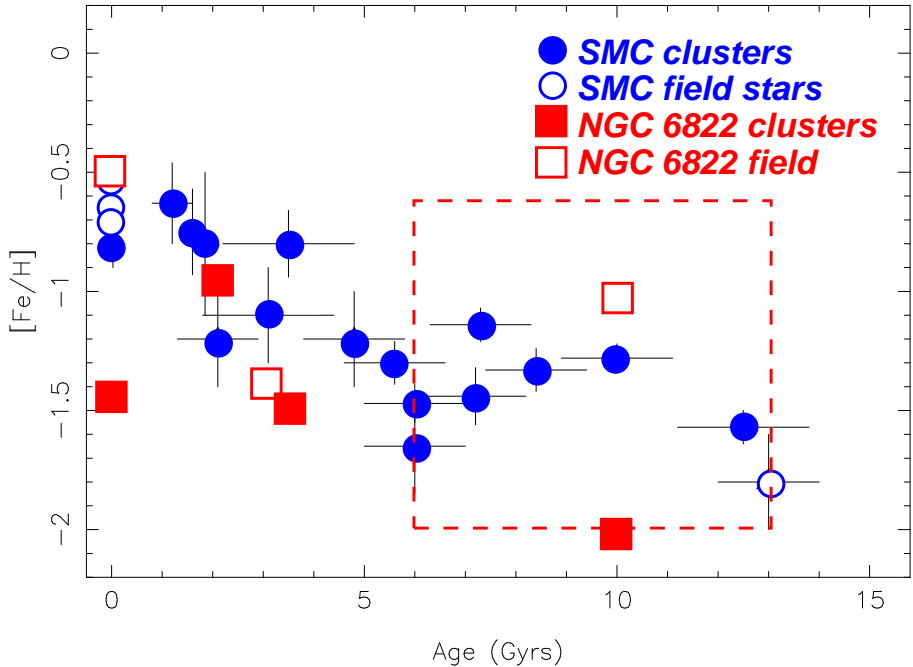}
\vskip0pt \caption{
Age versus metallicity for clusters and field stars in the 
SMC (circles) and in NGC 6822 (squares).  
The diagram for the
SMC was adopted from Da Costa (2002) and comprises both spectroscopic
and photometric abundances.  The data points for
NGC 6822 are based on a variety of different measurements and
methods (clusters: Cohen \& Blakeslee 1998; Chandar, Bianchi, \& Ford 2000;
Strader, Brodie, \& Huchra 2003; field: Muschielok et al.\ 1999; Tolstoy et 
al.\ 2001;  Venn et al.\ 2001) and should not be used to derive a quantitative 
age-metallicity relation.  The solid-line box denoting the mean
metallicity of NGC\,6822's red giant field population is shown for an assumed
age of 10 Gyr, while the much larger dashed box indicates the spread
in metallicity and gives a rough idea of the possible age range.
}
\label{fig3}
\end{figure*}

\subsection{Intermediate-age/Old Populations and Chemical Inhomogeneity}\label{Sect_AMR_old}

Whereas young populations in dIrrs tend to be fairly homogeneous,
old and inter\-mediate-age stellar populations show considerable
metallicity spreads.  In part this may be due to the large age
range sampled here, and the difficulty of assigning ages to individual
field stars.  Metallicity spreads have mainly been derived
based on the color width of the red giant branch in color-magnitude
diagrams, or via metallicity-sensitive photometric systems (e.g.,
Cole, Smecker-Hane, \& Gallagher 2000; Davidge 2003).  
These methods have the drawback that they are
affected by the age-metallicity degeneracy.  Near-infrared Ca\,{\sc ii}
triplet spectroscopy is now increasingly being employed for general
[Fe/H] derivations instead.
In the LMC, red giants in different parts of the galaxy
show significantly different mean abundances (Cole et al.\ 2000, 2004).
Cole et al.\ conclude that
in the LMC azimuthal metallicity variations may in part be due to
different fractions of bar and disk stars sampled at different
positions (with the bar stars being younger and more metal-rich).
With regard to field populations, star clusters have the advantage of
consisting of well-datable, single-age populations.
Old globular clusters in the LMC may differ substantially in metallicity
(Olszewski et al.\ 1991).  Interestingly, there is also evidence for a 
radial abundance gradient in the LMC old cluster population, i.e., 
a trend for old clusters to be more metal-rich closer to the LMC's
center (Da Costa 1999).
In the SMC, there are indications that intermediate-age star clusters
of a given age may occasionally differ by a few tenths of dex in [Fe/H]
(Da~Costa \& Hatzidimitriou 1998; Da~Costa 2002), which would 
indicate considerable chemical differences in the enrichment of
their birth clouds---possibly due to infall.  However, refined age
determinations and more spectroscopic abundance determinations are
needed to verify the SMC intermediate-age metallicity spread.

In NGC\,6822 and IC\,1613, abundance spreads among the field red giants
have been confirmed spectroscopically (Tolstoy et al.\ 
2001; Zucker \& Wyder 2004); again, age uncertainties remain.  
Although the present-day
field star metallicity in NGC\,6822 lies between those of the LMC and
the SMC, the cluster metallicities tend to lie below those of the 
SMC (Chandar, Bianchi, \& Ford 2000; Strader et al. 2003;
see Fig.\ \ref{fig3}).  One needs to caution that the cluster measurements
are based on a number of different studies and methods.  
Also, the present-day metallicity of NGC\,6822 may
have been enhanced by the recent interaction-triggered star formation
episode (see \S \ref{Sect_ISM_dwarfs}).

\subsection{Individual Element Abundances and Ratios}\label{Sect_AMR_elem}

The [$\alpha$/Fe] ratio in the Magellanic Clouds, NGC\,6822, and WLM
measured in the above studies is lower than the solar ratio.  There
are several possible reasons.  This may be a consequence of the lower
star formation rates in dIrrs or the possibility of a 
steeper initial mass function (see Tsujimoto et al.\ 1995; Pagel \&
Tautvai\v{s}ien\.e 1998), i.e., a reduced contribution of Type II supernovae 
as compared to Type Ia supernovae (e.g., Hill et al.\ 2000; Smith et al. 2002), 
%XX not I
or the possibility of metal loss through selective winds (Pilyugin 1996).
It seems that in the LMC, where $\alpha$ element abundances (most
notably oxygen) were measured in clusters of different ages, the
evolution of the [$\alpha$/Fe] ratio was fairly flat with time
(Hill et al.\ 2000 and Hill 2004).  Field red giants
also show reduced [O/Fe] values (Smith et al.\ 2002).  

The $r$-process 
(traced by, e.g.,  Eu), which is the dominant process in massive stars,
appears to prevail at low metallicities or the
early stages of the LMC's evolution.  The $s$-process 
(traced by, e.g., Ba and La), which takes place in cool
intermediate-mass giants such as asymptotic giant branch stars,
dominates at higher metallicities (Hill 2003).
Nitrogen appears to be close to primary and may come mainly from 
nonmassive stars (see also Maeder, Grebel, \& Mermilliod 1999).
Stellar and gaseous nitrogen abundances in the Magellanic Clouds show 
considerable variations.  Nitrogen enhancement in supergiants and old
red giants may be due to mixing of CN-enhanced material to their surface
during the first dredge-up (e.g., Russell \& Dopita 1992;
Venn 1999; Dufton et al.\ 2000; Smith et al.\ 2002;
not observed, however, in LMC B main sequence
stars; Korn et al.\ 2002).

\subsection{Age-metallicity Relations}\label{Sect_AMR_AMR}

In order to derive a reliable, detailed, age-metallicity relation suitable
to constrain the quantitative nature of the chemical evolution history of a 
galaxy, including the importance of infall, gas and metal loss, burstiness,
etc., one would ideally want very well-resolved temporal sampling.  
This is not yet possible with the currently available, sparse data.

Present-day abundances are traced well by H\,{\sc ii}
regions and massive stars.  Progress is being made at intermediate and old ages
using planetary nebulae and 
field red giants.  Planetary nebulae have recently been used for an
independent derivation of the age-metallicity evolution of the LMC at 
intermediate ages (Dopita et al.\ 1997).  While extragalactic planetary
nebulae cannot normally easily be age-dated, Dopita et al.\ extended their
spectroscopic data set for the LMC 
to the ultraviolet to try to directly measure the 
flux from the central star and also used the size information for the 
nebulae.  They were then able to not
only derive abundances but also ages using full photoionization modeling, 
and found their results in good 
agreement with stellar absorption-line spectroscopy.  
In less massive dIrrs, the number of planetary nebulae tends to be small,
making it more difficult to derive a well-sampled
age-metallicity relation.
In order to derive ages for individual
field stars, one has to complement the spectroscopic abundances by 
photometric luminosities and colors and rely on isochrone models.  
Considerably more accurate information can presently be obtained when using
star clusters as single-age, single-metallicity populations.  Disadvantages
of relying on star clusters are that one often only has very few such objects
in a galaxy, and that their properties are not necessarily 
representative of the field populations.

In all Irrs and dIrrs studied in some detail to date, there is evidence
for the expected increase in chemical enrichment with younger ages.  The LMC's
cluster age-metallicity relation clearly demonstrates this, although
there is the famous cluster age gap in the age range from $\sim 4$ to
$\sim 9$ Gyr (Da~Costa 2002, and references therein).  The SMC has the
unique advantage among all Local Group dIrrs to have formed and preserved
clusters throughout its lifetime.  While spectroscopic abundance determinations
and improved age determinations are still missing for many clusters, the
SMC shows a very well-defined age-metallicity relation with what appears
to be considerable metallicity scatter at a certain given age (see 
Fig.\ \ref{fig3}, adopted from Da~Costa 2002, and discussion in the
previous subsection).  While the age-metallicity relation appears to be
flatter than predicted by closed-box models in LMC and SMC,
Da~Costa (2002) notes that
the presently available data do not yet permit one to distinguish between
simple closed-box or leaky-box evolution models, bursty star formation 
histories, or infall.

Figure\ \ref{fig3} also shows data points for clusters and field
stars in NGC\,6822, for which rough age estimates and spectroscopic 
abundance determinations are available from a variety of different sources.  
As noted by Chandar et al.\
(2000), the clusters generally seem to be more metal-poor than the
SMC clusters and NGC\,6822's field population.  It is unclear whether
these differences would be reduced if all metallicities were determined 
using the same method, but overall the graph seems to indicate a trend
of increasing metallicity with decreasing age.  Undoubtedly, these kinds
of studies will be refined in the coming years.

\section{Other Global Correlations}\label{Sect_global}

\begin{figure*}
\centering
\includegraphics[width=0.77\columnwidth,angle=270]{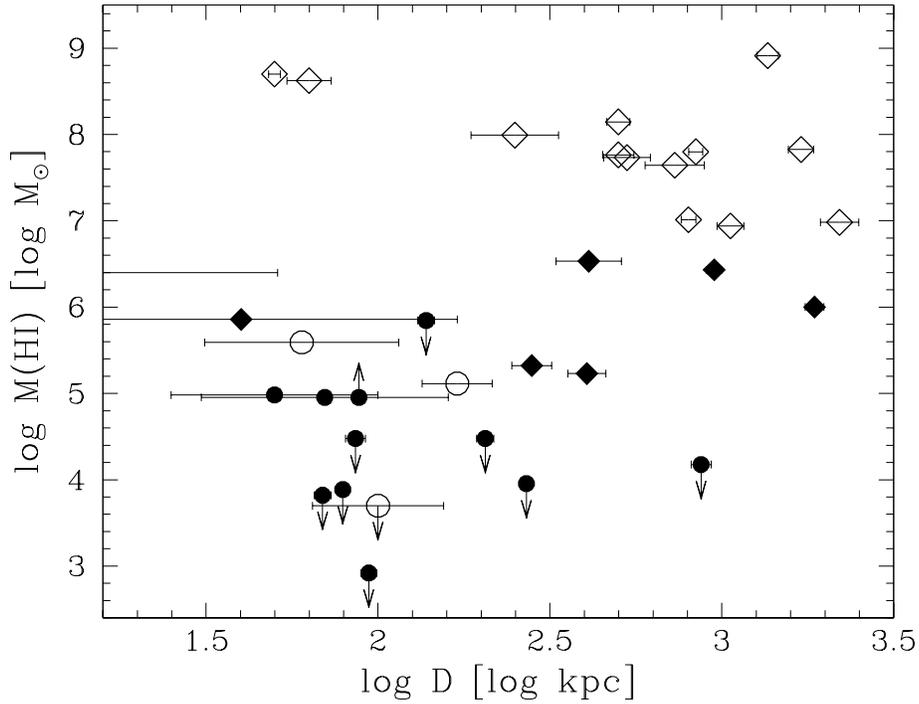}
\vskip0pt \caption{
Dwarf galaxy H\,{\sc i} mass versus distance to the nearest massive galaxy.
Filled circles stand for dwarf spheroidals (dSphs), open circles for dwarf
ellipticals, open diamonds for dwarf irregulars (dIrrs), and
filled diamonds for dIrr/dSph transition-type galaxies.
Lower or
upper H\,{\sc i} mass limits are indicated by arrows. There is a
general trend for the H\,{\sc i} masses to increase with increasing distance
from massive galaxies.  DSphs lie typically below $10^5$ $M_{\odot}$ in
H\,{\sc i} mass limits, while potential transition-type galaxies have
H\,{\sc i} masses of $\sim10^5$ to 10$^7$ $M_{\odot}$. DIrr galaxies
usually exceed  10$^7$ $M_{\odot}$.  (Figure from Grebel et al. 2003.)
}
\label{fig4}
\end{figure*}

\subsection{Gas and Environment}\label{Sect_global_gas}

When plotting galaxy H\,{\sc i} masses versus distance from the closest massive
galaxy in the Local Group and its immediate surroundings, 
we see a tendency for H\,{\sc i} masses
to increase with galactocentric distance (Fig.\ \ref{fig4}; Grebel et al.\
2003, and references therein).  Only fairly large
galaxies, such as the Magellanic Clouds, IC\,10, and
M33, with H\,{\sc i} masses $\gg 10^7$~
$M_{\odot}$, seem to be able to retain their gas reservoirs
when closer than $\sim 250$~kpc to giants (Grebel et al.\ 2003).
Note that weak lensing measurements and dynamical modeling indicate
typical dark matter halo scales for massive galaxies 
of $260\,h^{-1}$ kpc (e.g., McKay et al.\ 2002).  
%XX unit missing
The bulk of the Local Group dIrrs and transition-type galaxies are
located beyond $\sim$250~kpc from M31 and the Milky Way.
At these distances these gas-rich galaxies seem to be less
prone to galaxy harassment (i.e., in this case loss of gas through
interactions with spirals), although the details will depend on their
(yet unknown) orbital parameters (Grebel et al.\ 2003).
A similar trend for dIrrs and dIrr/dSphs is seen in the Sculptor group
(Skillman, C\^ot\'e, \& Miller 2003a).  

Skillman, C\^ot\'e, \& Miller (2003b) investigated correlations between
H\,{\sc i} mass fraction and metallicity (oxygen abundance) for Local
Group dIrrs and dIrrs in the Sculptor group of galaxies.  They found
that the Local Group dIrrs deviate from model curves expected for closed-box
evolution, whereas the Sculptor group dIrrs follow these curves rather
closely.  Indeed the dIrrs that exhibit the strongest deviations are those
with very low metallicity and gas content.  Considering that the Sculptor
group is, in contrast to the Local Group, a very loose and diffuse group
or cloud (Karachentsev et al.\ 2003b), this may indicate that closed-box
evolution is more likely to occur in low-density environments with little
harassment.  

\subsection{Luminosity and Metallicity}\label{Sect_global_LZ}

\begin{figure*}
\centering
\includegraphics[width=0.77\columnwidth,angle=270]{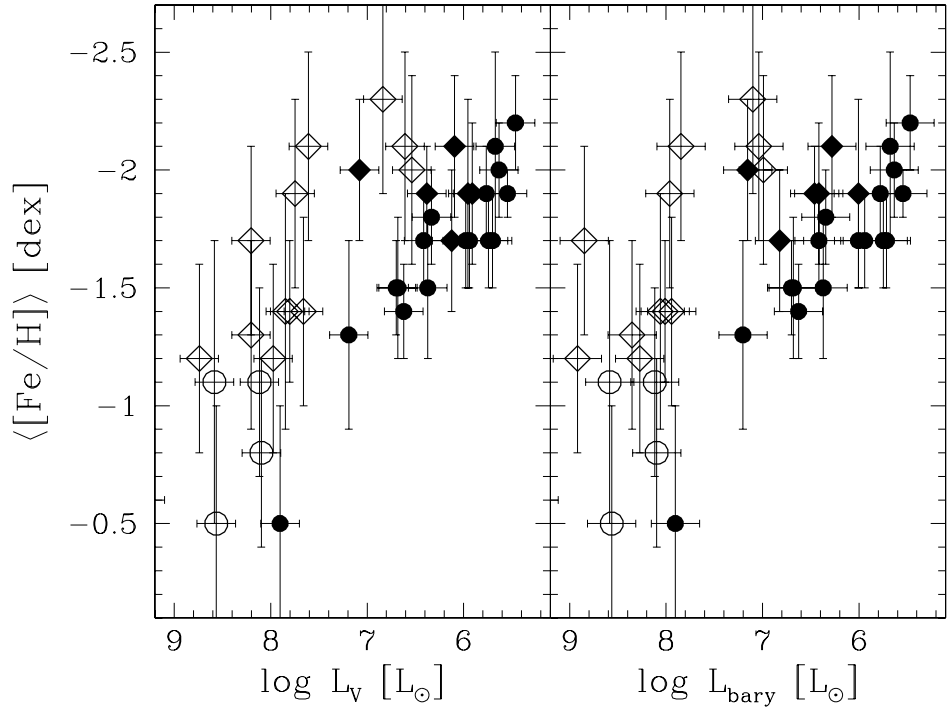}
\vskip0pt \caption{
$V$-band luminosity ({\it left panel}) and baryonic luminosity ({\it right
panel}, corrected for baryon contribution of gas not yet turned into stars)
versus mean metallicity of red giants.  The symbols are the same as in
Fig.\ \ref{fig4}.  The error bars in metallicity indicate the metallicity
spread in the old populations, not the uncertainty of the metallicity.
DIrrs are more luminous at equal metallicity than dSphs.  
Or, in other words, dSphs are too metal rich for their low luminosity.
However, several
dIrr/dSph transition-type galaxies coincide with the dSph locus.  These
objects are indistinguishable from dSphs in all their properties except
for gas content.  (Figure from Grebel et al. 2003.)
}
\label{fig5}
\end{figure*}

Mean galaxy metallicity and mean galaxy luminosity are well correlated,
as has been known for a long time.  For dIrrs, one usually considers
present-day oxygen abundances, and when comparing them to other galaxy
types without ionized gas (such as dSphs), stellar [Fe/H] values are
converted into what is assumed to be the corresponding nebular abundance.
This conversion comes with a number of uncertainties.  Grebel et al.\
(2003) therefore used the stellar (red giant) metallicities of Local Group 
dwarf galaxies of all types to directly compare the properties of their
old populations.  A plot of $V$-band luminosity $L_V$ versus
$\langle$[Fe/H]$\rangle$ (Fig.\ \ref{fig5}, left panel)
shows a clear trend of increasing luminosity with
increasing mean red giant branch
metallicity.  However, different galaxy types (gas-rich dIrrs and
gas-deficient dSphs) are offset from each other in that the dIrrs 
are more luminous at the same
metallicity.

In other words, the dIrrs have too low a metallicity for their
luminosity as compared to dSphs.  {\em Thus, dSphs, most of which have been
quiescent over at least the past few Gyr, must have experienced chemical
enrichment faster and more efficiently than dIrrs, which continue to
form stars until the present day} (Grebel et al.\ 2003).  It is tempting
to speculate that environment may once again have affected this in the 
sense of a denser environment leading to more vigorous early
star formation rates.

When plotting the baryonic luminosity (Milgrom \& Braun 1988;
Matthews, van~Driel, \& Gallagher 1998)
against metallicity (Fig.\ \ref{fig5},
right panel), the locus of
the dSphs remains unchanged while the dIrrs move to
higher luminosities as compared to the dSphs.
Thus, if star formation in present-day dIrrs were terminated when all of
their gas was converted into stars, then these fading dIrrs would be even
further from the dSph luminosity-metallicity relation.
For a discussion of the amount of fading, time scales, angular momentum
loss, etc. required for converting a dIrr into a dSph, see Grebel et al.\
(2003).  Here we simply want to emphasize that dIrrs follow a 
metallicity-luminosity relation that requires a different evolutionary
path than in other types of dwarf galaxies.  In particular, it seems that
dIrrs are an intrinsically different type of galaxy than dSphs.  We note
in passing that for dIrrs not only do metallicities correlate well with
luminosities, but also with surface brightness.

\section{Summary}\label{Sect_Summ}

Irr and dIrr galaxies are usually gas-rich galaxies with ongoing or
recent star formation.  They are preferentially found in the outer
regions of groups and clusters as well as in the field.  Irrs and
massive dIrrs exhibit solid body rotation, while low-mass dIrrs seem
to be dominated by random motions.  Spiral density waves are absent.

Irrs and dIrrs are often embedded in extended H\,{\sc i} halos, which,
in the absence of interactions, appear fairly regular.  In low-mass
dIrrs, the centroid of the H\,{\sc i} distribution does not necessarily
coincide with the optical center of the galaxy, and occasionally annular
structures are seen.  The neutral gas tends to be flocculent, dominated
by shells and bubbles, and driven by the turbulent energy input from
massive stars and supernovae.  Molecular gas and dust form less easily
and are more easily dissociated due to the high UV radiation field
and fewer coolants in low-metallicity environments.  

Irrs and dIrrs
usually contain multiple distinct zones of concurrent star formation.
Extended regions of active star formation tend to be long-lived and
gradually migrate on time scales on a few tens to hundreds of Myr.  
Stochastic self-propagating star formation seems to be the main driver
of star formation activity.  There is no need for external triggering.
In quiescently evolving dIrrs and/or dIrrs with slow or no rotation
(usually the less massive dIrrs), 
the degree of central concentration of star formation is small, while
the reverse trend is true for more massive and faster rotators.  The
formation of populous clusters seems to be preferred in more massive
and/or interacting dIrrs.  Generally, gas consumption is sufficiently
low that star formation in Irrs and dIrrs may continue for another
Hubble time.  On global scales the star formation rate of Irrs and dIrrs
is close to constant, with amplitude variations of factors of 2--3.

Old stellar populations are ubiquitous in all Irrs and dIrrs studied in detail 
so far, although their fractions vary widely.  
In contrast to the many scattered young OB associations and
superassociations, older populations show a smooth and regular
distribution that is much more extended than that of the young
populations.  Both young stellar populations and H\,{\sc ii} regions
agree very well in their abundances, underlining the chemical homogeneity
of Irrs and dIrrs.  However, taken at face value, intermediate-age and
old populations tend to exhibit considerably more scatter in their 
metallicity.  There are indications that star clusters of the same age
may differ by several tenths of dex in metallicity, although observational
biases cannot yet be fully ruled out.  Overall, Irrs and
dIrrs follow the expected trend of increasing metal enrichment toward
younger ages; the currently available data do not yet permit one to
unambiguously distinguish between infall and leaky-box
versus closed-box chemical evolution, nor to reliably evaluate the 
importance and impact of possible bursts.

Substantial
progress is being made not only in spectroscopic measurements of stellar
metallicities, but also in the determination of individual element 
abundances.  The [$\alpha$/Fe] ratios in Irrs and dIrrs, which tend to
be lower than the solar ratio, and the lower effective yields
may be interpreted as indicative of lower
astration rates and a reduced contribution of Type II supernovae.
Other interpretations (different initial mass functions, leaky-box chemical 
evolution
with metal loss through selective winds) are being entertained as well.

Correlations between gas content and distance from massive galaxies 
as well as morphological segregation 
indicate that environment (in particular gas loss through 
ram pressure or tidal stripping; see also Parodi, Barazza, \& Binggeli
2002; Lee, McCall, \& Richer 2003
for the Virgo cluster) does have an impact on the evolution of
Irrs and dIrrs.  Irrs and dIrrs follow the well-known relation of
increasing mean metallicity with increasing galaxy luminosity.  The offset
in this relation from the relation for dSphs, such that dIrrs are more
luminous than dSphs at the same metallicity, indicates that the early
chemical evolution in these two galaxy types proceeded differently, with
dSphs becoming enriched more quickly.

\vspace{0.3cm}
{\bf Acknowledgements}.
I would like to thank the organizers, particularly Andy McWilliam, 
for their kind invitation to this very interesting Symposium, and for their 
patience while my paper was finished.  I am grateful to Jay Gallagher for a 
critical reading of the text.

\end{document}